\documentclass[12pt]{iopart}
\usepackage{amssymb}
\usepackage{graphicx}
\usepackage{epstopdf}
\usepackage{hyphenat}
\usepackage{multirow}
\usepackage{subfig}
\usepackage{upgreek}
\usepackage{float}

\begin{document}

\newcommand{\summ}{\displaystyle\sum}

\title[Flux line relaxation following current quenches]{Flux line relaxation kinetics following current quenches in disordered type-II superconductors}

\author{Harshwardhan Chaturvedi$^1$, Hiba Assi$^1$, Ulrich Dobramysl$^2$, Michel Pleimling$^{1,3}$, and Uwe C. T\"auber$^1$}

\address{$^1$ Department of Physics \& Center for Soft Matter and Biological Physics, Virginia Tech, Blacksburg, Virginia 24061-0435, United States}
\address{$^2$ Wellcome Trust / CRUK Gurdon Institute, University of Cambridge, Tennis Court Rd, Cambridge CB2 1QN, United Kingdom}
\address{$^3$ Academy of Integrated Science, Virginia Tech, Blacksburg, Virginia 24061-0405, United States}

\begin{abstract}
We investigate the relaxation dynamics of magnetic vortex lines in type-II superconductors following rapid changes of the external driving current by means of an elastic line model simulated with Langevin molecular dynamics. A system of flux vortices in a sample with randomly distributed point-like defects is subjected to an external current of appropriate strength for a sufficient period of time so as to be in a moving non-equilibrium steady state. The current is then instantaneously lowered to a value that pertains to either the moving or pinned regime. The ensuing relaxation of the flux lines is studied via one-time observables such as their mean velocity and radius of gyration. We have in addition measured the two-time flux line height autocorrelation function to investigate dynamical scaling and aging behavior in the system, which in particular emerge after quenches into the glassy pinned state.
\end{abstract}
\pacs{87.23.Cc, 02.50.Ey, 05.40.-a, 87.18.Tt}

\submitto{\JSTAT --- \today}

\section{Introduction}\label{intro}
Driven vortex lines in type-II superconductors, in the presence of randomly distributed point-like quenched disorder, constitute a rich, highly complex system far from equilibrium, with a number of competing energy, time and length scales. This system displays a variety of thermodynamic phases and intriguing transport properties \cite{Blatter1994}, making it interesting from a statistical physics viewpoint. When driven by Lorentz forces generated by an external current, these magnetic flux vortices move through the superconducting sample generating an electric field that opposes the external current, resulting in Ohmic dissipation. Quenched disorder in the form of material defects that effectively act as attractive pinning centers can be optimally distributed in the system to curb flux flow, consequently restoring dissipation-free transport \cite{Blatter1994}, the property that renders these materials desirable for technology (such as generation of strong magnetic fields for MRI scanners and particle accelerators).

Uncorrelated point-like disorder can be naturally occurring (e.g. oxygen vacancies) or artificially introduced (e.g. by electron irradiation \cite{Kwok1994}). Weak uncorrelated disorder is known to destroy the long-range translational order of the Abrikosov flux line lattice that forms at low temperatures in a three-dimensional system free of disorder, replacing it with either a vortex glass phase completely devoid of translational order \cite{Fisher1989,Feigelman1989,Nattermann1990,Fisher1991,Kwok1992} or a Bragg glass phase with quasi long-range positional order \cite{Giamarchi1994,Giamarchi1995,Kierfeld1997,Fisher1997,Giamarchi1997,Nattermann2000}.

An applied external electric current exerts a transverse Lorentz force on the flux lines. At ambient temperatures much lower than the temperature at which the vortex glass melts into a flux liquid in the absence of an external current, the introduction of external current transforms the disorder-dominated equilibrium phase into a creeping Bragg glass state. As the current is increased beyond the critical depinning transition value, the stresses in the flux line lattice increase sufficiently enough to result in the breaking of the lattice into a moving liquid. At even larger drive, transverse order develops in the liquid to result in the formation of a moving smectic which when subjected to even higher drive dynamically freezes into a moving Abrikosov flux lattice \cite{Nattermann2000}. In our numerical simulations, finite-size effects resulting from the small system size and low number of flux lines render it difficult for us to distinguish between the moving liquid, smectic, and moving lattice states. We therefore limit ourselves in the present work to a broad classification of states of vortex matter into three regimes of current viz. the {\it moving regime} corresponding to {\it high} current values, the {\it pinned regime} corresponding to {\it low} current values and the {\it critical regime} associated with {\it intermediate} values; the terms {\it high}, {\it low} and {\it intermediate} are quantified in Section \ref{secRegimes} below.

A material shows physical \lq aging' when it undergoes slow relaxation from a state away from equilibrium to reach thermal equilibrium and displays breaking of time-translation invariance in this non-stationary regime \cite{Struik1978,Henkel2010}. Aging is often seen in frustrated environments characterized by a large number of energetically close metastable states, that is, in glassy systems \cite{Henkel2007}. The discovery of the dependence of the voltage response to an applied current in superconducting 2H-NbSe$_2$ on the duration of the application clearly indicated the presence of physical aging in disordered vortex matter \cite{Du2007}. Bustingorry, Cugliandolo, and Dom\'inguez studied a three-dimensional elastic line model of vortex matter by means of Langevin molecular dynamics (LMD) to identify physical aging features in two-time quantities like  density-density autocorrelation and mean-square displacement \cite{Bustingorry2006,Bustingorry2007}. The aforementioned studies utilized a random landscape representation for the distribution of disorder while we implement isolated localized pinning centers of uniform potential, two approaches that yield significantly different relaxation properties \cite{Dobramysl2014} and aging scaling exponents \cite{Pleimling2011}. Pleimling and T\"auber investigated the non-equilibrium relaxation properties of vortex matter as elastic lines by means of Monte Carlo simulations \cite{Pleimling2011}; Dobramysl {\em et al.} later verified these results with a different microscopic representation of the system's dynamics through LMD \cite{Dobramysl2013}. Assi {\em et al.} proceeded to use the latter LMD implementation of the system to study relaxation dynamics of vortex lines following magnetic field and temperature quenches to analyze the system's sensitivity to sudden external perturbations \cite{Assi2015}.

In the present work, we employ Langevin molecular dynamics to study the aging relaxation dynamics of flux lines in type-II superconductors when subjected to sudden changes in the externally applied current. We accomplish this by subjecting our coarse-grained elastic lines representing vortex matter to an instantaneous change in driving force after letting the lines relax under the influence of the initial drive sufficiently long for them to have reached a moving, non-equilibrium steady state. We either quench the current within the moving regime or from the moving into the pinned regime. We perform these simulations first with repulsive vortex interactions turned off, and subsequently switched on, in order to identify the physical mechanisms behind the complex features seen in the vortex relaxation dynamics. We investigate this rich non-equilibrium relaxation behavior by the measurement of various one- and two-time observables.

The organization of this paper is as follows. The next section explains the elastic line model, defines the LMD algorithm we employ to implement its stochastic dynamics, and specifies the material parameters we use for the implementation. We then describe the simulation protocol for the drive quenches and define the one- and two-time observables we measure in order to quantify the relaxation behavior of the system post-quench. We end the section by quantifying the pinned and moving regimes, based on steady-state results for the one-time quantities measured in the system. Section 3 is devoted to results obtained from studying the relaxation properties of systems of flux lines undergoing sudden current changes. We compare the effects of instantaneously decreasing or down-quenching the drive between values in the moving regime to results for quenches from the moving to the pinned regime, which yield markedly different behavior. We also isolate the effect of inter-vortex interactions on the relaxation phenomena by alternately running the simulations with interactions switched off and on. We conclude the paper by summarizing our results in Section 4.

\section{Elastic Line Model and Simulation Protocol}\label{sec2}

\subsection{Model Hamiltonian}\label{secHam}

We model magnetic flux lines dynamically in the extreme London limit (where the London penetration depth is much larger than the coherence length) as mutually repulsive elastic lines \cite{Nelson1993,Das2003}. We write down the Hamiltonian of the system as a sum of three competing terms viz. the elastic line tension energy, the attractive potential due to point-like pinning sites, and the mutual repulsive interactions between flux lines:
\begin{equation}\label{Ham}
\fl H[\mathbf{r}_{i,z}(t)]=\summ_{i=1}^{N}\int_{0}^{L}dz \Bigg[ \frac{\tilde{\epsilon}_1}{2}\left\vert\frac{d\mathbf{r}_{i,z}(t)}{dz}\right\vert^2+U_D(\mathbf{r}_{i,z}(t),z) + \frac{1}{2}\summ_{j\neq i}^{N}V(|\mathbf{r}_{i,z}(t) - \mathbf{r}_{j,z}(t)|)\Bigg].
\end{equation}
$\mathbf{r}_{i,z}(t)$ is the position vector in the $xy$-plane at time $t$, of the line element of the $i$th flux line (one of $N$), at height $z$ in the vertical direction, which is also the direction of the applied external magnetic field. The elastic line stiffness or local tilt modulus is given by $\tilde{\epsilon}_1 \approx \Gamma^{-2}\epsilon_0\ln(\lambda_{ab}/\xi_{ab})$ where $\Gamma^{-1}=M_{ab}/M_c$ is the effective mass ratio or anisotropy parameter. $\lambda_{ab}$ is the London penetration depth and $\xi_{ab}$ is the coherence length, in the $ab$ crystallographic plane. The in-plane repulsive interaction between any two flux lines is given by $V(r)=2\epsilon_0K_0(r/\lambda_{ab})$, where $K_0$ denotes the zeroth-order modified Bessel function. It effectively serves as a logarithmic repulsion that is exponentially screened at the scale $\lambda_{ab}$. The pinning sites are modeled as randomly distributed smooth potential wells, given by
\begin{equation}\label{pin_pot}
U_D(\mathbf{r}, z)=-\summ_{\alpha=1}^{N_D}\frac{b_0}{2}p\left[1-\tanh\left(5\frac{|\mathbf{r}-\mathbf{r}_\alpha|-b_0}{b_0}\right)\right]\delta(z-z_\alpha),
\end{equation}
where $N_D$ is the number of pinning sites, $p\ge0$ is the pinning potential strength, and $\mathbf{r}_\alpha$ and $z_\alpha$ respectively represent the in-plane and vertical position of pinning site $\alpha$. In the following, all lengths are measured in units of the pinning potential width $b_0$. Energies are measured in units of $\epsilon_0 b_0$, where $\epsilon_0=(\phi_0/4\pi\lambda_{ab})^2$ is the elastic line energy per unit length, and  $\phi_0=hc/2e$ is the magnetic flux quantum.

\subsection{Langevin Molecular Dynamics}\label{secLan}

In order to simulate the dynamics of the model, we discretize the system along the $z$ axis, i.e., the direction of the external magnetic field, into layers, with the layer spacing corresponding to the crystal unit cell size $c_0$ along the crystallographic $c$ direction \cite{Das2003,Bullard2008}. Consequently, each elastic line is broken up into points, with each point belonging to a given line, residing in a unique layer. Any two points of the same line in neighboring layers attract each other via an elastic force, the potential between them constituting the first term in the Hamiltonian \eref{Ham}. Points in the same layer repel each other via long-range logarithmic interactions that are defined by the third term of the Hamiltonian. The pinning sites are also confined to these layers perpendicular to the $z$ axis, and are modeled as smooth potential wells \eref{pin_pot}. The interactions between the discrete elements of the system that are described here are encapsulated in the properly discretized version of the Hamiltonian. We use this discretized Hamiltonian to obtain coupled, overdamped Langevin equations which we solve numerically:
\begin{equation}\label{lan}
\eta\frac{\partial\mathbf{r}_{i,z}(t)}{\partial t}=-\frac{\delta H[\mathbf{r}_{i,z}(t)]}{\delta\mathbf{r}_{i,z}(t)}+\mathbf{f}_{i,z}(t).
\end{equation}
Here, $\eta=\phi_0^2/2\pi\rho_n c^2 \xi_{ab}^2$ is the Bardeen-Stephen viscous drag parameter, where $\rho_n$ represents the normal-state resistivity of YBCO near $T_C$ \cite{Blatter1994,Bardeen1965}. We model the fast, microscopic degrees of freedom of the surrounding medium by means of thermal stochastic forcing as uncorrelated Gaussian white noise $\mathbf{f}_{i,z}(t)$ with vanishing mean $\langle\mathbf{f}_{i,z}(t)\rangle=0$. Furthermore, these stochastic forces obey the Einstein relation 
\begin{equation}\nonumber
\langle\mathbf{f}_{i,z}(t) \cdot \mathbf{f}_{j,z'}(s)\rangle = 4\eta k_BT\delta_{ij}\delta_{zz'}\delta(t-s),
\end{equation}
which ensures that the system relaxes to thermal equilibrium with a canonical probability distribution $P[\mathbf{r}_{i,z}]\propto e^{-H[\mathbf{r}_{i,z}]/k_BT}$ in the absence of external current.

\subsection{Model Parameters}\label{secParam}
We have selected our model parameters to closely match the material properties of the ceramic high-$T_C$ type-II superconductor YBa$_2$Cu$_3$O$_7$ (YBCO). The pinning center radius is set to $b_0=35\AA$; all simulation distances are measured in units of this quantity. The inter-layer spacing in the crystallographic $c$ direction is set to this microscopic scale, $c_0=b_0$. The in-plane London penetration depth and superconducting coherence length are chosen to be $\lambda_{ab}=34b_0\approx 1200\AA$ and $\xi_{ab}=0.3b_0\approx 10.5\AA$ respectively, in order to model the high anisotropy of YBCO, which has an effective mass anisotropy ratio $\Gamma^{-1}=1/5$. The line energy per unit length is $\epsilon_0\approx1.92\cdot 10^{-6}\mathrm{erg}/\mathrm{cm}$; all simulation energies are measured in units of $\epsilon_0 b_0$. This effectively renders the vortex line tension energy scale to be $\tilde{\epsilon}_1/\epsilon_0\approx 0.189$. The pinning potential well depth is set to $p/\epsilon_0=0.05$. The temperature in our simulations is set to $10\,$K ($k_{\rm B} T / \epsilon_0 b_0=0.002$ in our simulation units). The Bardeen-Stephen viscous drag coefficient $\eta=\phi_0^2/2\pi\rho_nc^2\xi_{ab}^2 \approx10^{-10} \, \mathrm{erg} \cdot\mathrm{s} / \mathrm{cm}^2$ is set to one, where $\rho_n \approx 500 \, \mathrm{\mu \Omega m}$ is the normal-state resistivity of YBCO near $T_C$ \cite{Abdeladi1994}. This results in the simulation time step being defined by the fundamental temporal unit $t_0=\eta b_0/\epsilon_0\approx 18\,$ps; all times are measured in units of $t_0$.

\subsection{Drive Quench Simulation Protocol}\label{secProt}

Our system consists of $N=16$ flux lines, moving in a three-dimensional space with periodic boundary conditions in the $xy$ directions and free boundary conditions along the $z$ direction. The system is discretized into $L=100$ layers along the $z$ direction. We simulate point-like disorder by randomly distributing $1116$ pinning sites per layer, using a different random distribution for each layer. We set the horizontal system size to $(16/\sqrt{3}\lambda_{ab}\times 8\lambda_{ab})$. This ratio of horizontal boundary lengths is necessary to ensure that the system, in the absence of disorder or drive, equilibrates to a state where the flux lines arrange themselves into a periodic hexagonal Abrikosov lattice.

Each simulation run starts with perfectly straight flux lines, distributed randomly in the computational space. The Lorentz force exerted on the flux lines by an external current is modeled in the system as a tunable, spatially uniform drive $F_d$ in the $x$ direction, the introduction of which requires the addition of a work term $-\mathbf{F_d}\cdot\mathbf{r}_{i,z}(t)$ to the Hamiltonian \eref{Ham} and hence $\mathbf{F_d}$ to the right-hand side of the Langevin equations \eref{lan}. Having set $F_d$ to some initial value, the lines are left to relax beyond microscopic time scales in a temperature bath at $T=0.002$ in our dimensionless units, or $10$ K. The strength of the initial drive is chosen according to whether we want to start with a system in a moving or pinned state; we discuss these states in detail in Section \ref{secRegimes}. During this time, thermal fluctuations contribute towards the roughening of the lines. After this initial relaxation period of $60,000t_0$, we instantaneously change the drive to a different value that once again corresponds to a state in either the pinned or moving regime. At this point, we reset the system clock $t$ to $0$. All physical quantities are measured with respect to $t$. Following the drive quench, we start the measurement of one-time quantities and allow the system to relax for waiting time $s$. After the waiting time has elapsed, we take a snapshot of the system. We then begin the measurement of two-time quantities which continues until the end of the simulation, with a run time of $500,000 t_0$ after the drive quench (set at $t=0$).

\subsection{Measured Quantities}\label{mesQua}
In the course of a simulation, we measure one- and two-time physical quantities, all of which are averaged over many disorder realizations and noise histories. A one-time quantity of interest is the mean $velocity$ of the lines, measured by averaging the velocity of all line elements at time $t$:
\begin{equation}
\mathbf{v}(t) = \left\langle \frac{d}{dt}\mathbf{r}_{i,z}(t)  \right\rangle_{z,i}.
\end{equation}
$\langle \ldots \rangle_z$ represents an average over the $z$ axis, i.e., over all vertical layers, while $\langle \ldots \rangle_i$ represents an average over all the lines. Another one-time quantity we measure is the mean \textit{radius of gyration}
\begin{equation}\label{GR}
r_g(t)=\sqrt{\langle(\mathbf{r}_{i,z}(t)-\langle\mathbf{r}_{i,z}(t)\rangle_z)^2\rangle_{z,i}}\ .
\end{equation}
The radius of gyration is defined as the standard deviation of the lateral positions $\mathbf{r}_{i,z}(t)$ of the points constituting the $i$th flux line, averaged over all the lines. $r_g(t)$ is a measure of the roughness of the lines in the system, which is produced by thermal spatial fluctuations and line distortion due to pinning of line elements by pinning centers distributed in the sample. The third one-time quantity we measure is the \textit{fraction of pinned line elements}, i.e.,
\begin{equation}\label{GR1}
f_p = N(r<b_0)/N.
\end{equation}
It is defined as the fraction of line elements in the system that are located within distance $b_0$ (pinning center radius) of a pinning center.

The two-time quantity we measure in this study is the normalized \textit{height autocorrelation} function
\begin{equation}
C(t,s)=\frac{\left<(\mathbf{r}_{i,z}(t)-\langle\mathbf{r}_{i,z}(t)\rangle_z)(\mathbf{r}_{i,z}(s)-\langle\mathbf{r}_{i,z}(s)\rangle_z)\right>_{z,i}}{\left<(\mathbf{r}_{i,z}(s)-\langle\mathbf{r}_{i,z}(s)\rangle_z)^2\right>_{z,i}}\ .
\end{equation}
It quantifies how the lateral positions $\mathbf{r}_{i,z}$ of the elements of a line relative to the mean lateral line position $\langle\mathbf{r}_{i,z}\rangle_z$ at the present time $t$ are correlated to those relative positions at a past time $s$ and contains information about local transverse thermal fluctuations of vortex line elements. This quantity is averaged over all lines as well as over several thousand noise histories and disorder realizations. It is worth noting that the term `height' autocorrelation originates from viewing the flux lines as fluctuating one-dimensional interfaces, the local height of which corresponds to the deviation of $\mathbf{r}_{i,z}$ from the respective line's mean position. We use the height autocorrelations as a tool to investigate the existence and nature of physical aging in our system. A system shows aging when a dynamical two-time quantitiy displays slow relaxation and the breaking of time translation invariance  \cite{Henkel2010}. Additionally, in a {\it simple aging} scenario, the two-time quantity shows dynamical scaling and follows the general scaling form 
\begin{equation}
C(t,s)=s^{-b}f_C(t/s),
\end{equation}
where $f_C$ is a scaling function that follows the asymptotic power law 
\begin{equation}\label{scaform}
f_C(t/s)\sim(t/s)^{-\lambda_C/z},
\end{equation}
as $t\rightarrow\infty$; $b$ is the aging scaling exponent, $\lambda_C$ is the autocorrelation exponent, and $z$ is the dynamical scaling exponent.

\subsection{Moving and Pinned Regimes}\label{secRegimes}

\begin{figure}
\centering
  \subfloat{\includegraphics[width=.97\linewidth]{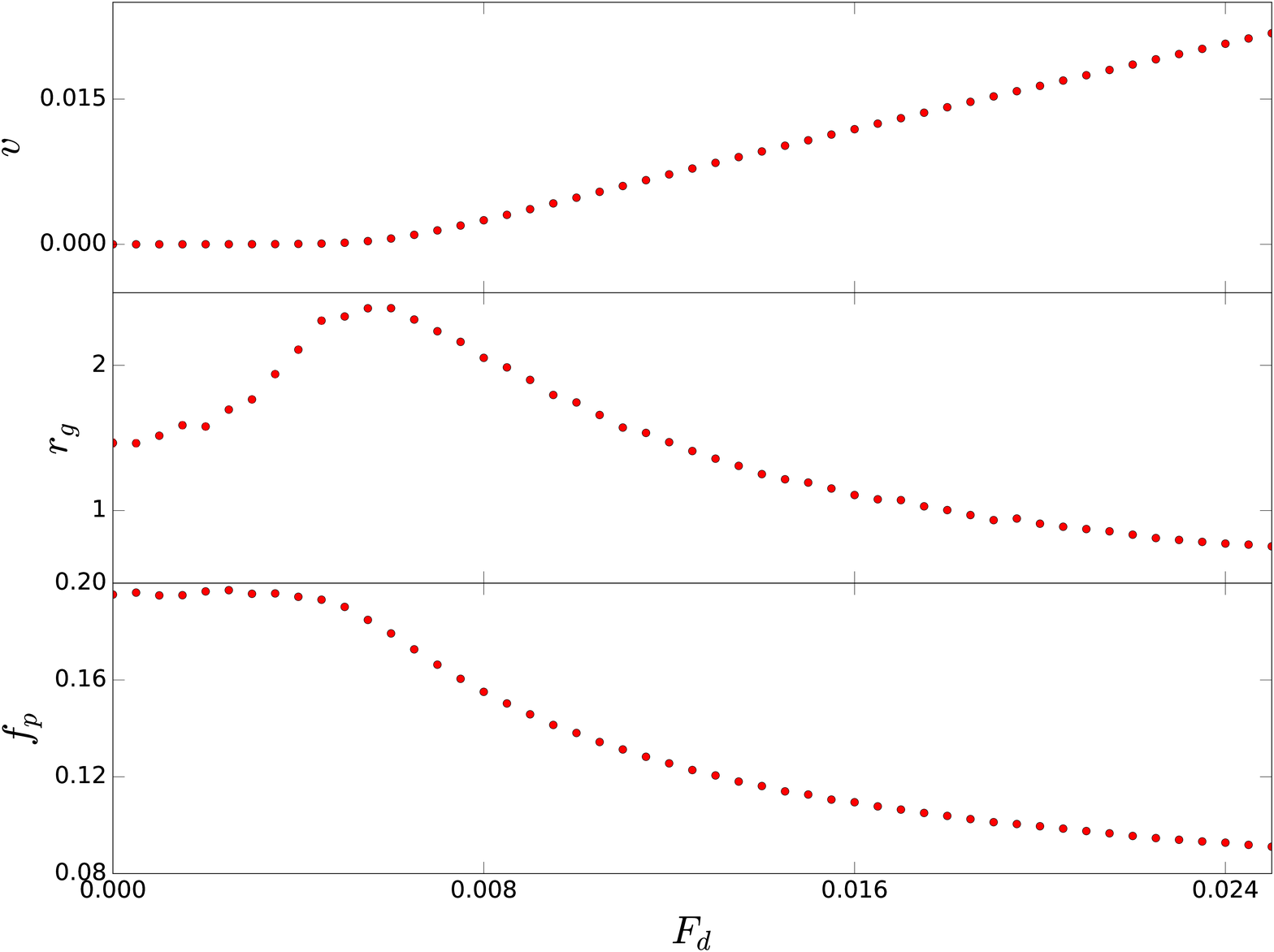}
  \llap{\parbox[b]{5.4in}{{\Large $(a)$}\\\rule{0ex}{4.05in}}}
  \llap{\parbox[b]{5.45in}{{\Large $(b)$}\\\rule{0ex}{2.65in}}}
  \llap{\parbox[b]{5.5in}{{\Large $(c)$}\\\rule{0ex}{1.35in}}}}
  \caption{Steady-state (a) mean vortex velocity $v$ (units of $b_0/t_0$), (b) radius of gyration $r_g$ (units of $b_0$), and (c) fraction of pinned line elements $f_p$ as a function of drive $F_d$ (units of $\epsilon_0$) for a system of interacting flux lines. $r_g$ peaks at $F_d \approx 0.006\epsilon_0$, where also $v$ starts assuming non-zero values, and $f_p$ begins to decay from its pinned steady state value $\sim 0.2$. Data are averaged over 100 realizations. \label{steady_st}}
\end{figure}

In order to identify the drive ranges that correspond to states when the system of flux lines is respectively in the pinned or moving regime, we have investigated steady-state features of the system as a function of drive, in the following manner: For a system of interacting flux lines, we set the drive to a certain value and allowed the system to evolve for $60,000t_0$ to let the system arrive to a non-equilibrium steady state. At this point, we started measuring the one-time physical quantities of the system viz. the radius of gyration $r_g$ and velocity $v$ at intervals of $t_0$ for $250,000t_0$. We performed this operation for $50$ evenly spaced drive values between $F_d = 0$ and $0.025\epsilon_0$. We averaged our results over the final $250,000$ time-steps and over 100 independent realizations, effectively averaging over $25$ million values per $F_d$ value. Error bars, representing the statistical error or standard deviation of the mean obtained via the aforementioned averaging process, are smaller than the symbol sizes in Fig. \ref{steady_st}. Similarly, the data points in every figure appearing in this paper represent mean values obtained by averaging over several independent realizations (the exact number of realizations are specified in the figure captions) and are accompanied by error bars representing the standard deviation of the mean in case these error bars are larger than the symbol sizes.

For zero drive, about $20\%$ (see $f_p$ in Fig. \ref{steady_st}c) of flux line elements are pinned by the pinning centers, as they have had $60,000$ time steps to move around the system exclusively via thermal wandering and find point-like defects that will trap them. The absence of drive further increases the likelihood of the flux lines remaining relatively motionless and trapped in their pinned configurations, as seen by the zero mean velocity of the lines at $F_d = 0$ (Fig. \ref{steady_st}a). Upon introducing drive, at small values, we see an increased radius of gyration compared to the case with $F_d = 0$. This can be attributed to the relatively weak drive assisted by thermal fluctuations causing portions of the lines that are weakly pinned to break free from their original pins and get trapped in other nearby pins resulting in distortions of the line configurations which translates to increased line roughness and hence larger gyration radius $r_g$. The persistence of the pinned state under these drive conditions is supported by the continued absence of mean line velocity $v$ and the lack of significant change in the fraction of pinned line elements $f_p$ compared to its value ($\approx0.20$) at $t=0$. The radius of gyration continues to increase with drive, until the drive is large enough to overcome the attractive forces exerted by the pins, enabling a complete depinning of the lines from the pins. This depinning point is marked by the rise of $v$, coinciding with a drop in $r_g$ and $f_p$. These trends continue for the remainder of the drive values, resulting in the flux lines getting further depinned (lower $f_p$), moving faster (higher $v$) and becoming straighter (lower $r_g$) with increasing drive. The depinning crossover appears to occur somewhere in the drive interval $0.004 \epsilon_0 \leq F_d \leq 0.008 \epsilon_0$, the \textit{critical regime} of drive. Drive values below this interval ($F_d<0.004\epsilon_0$) constitute the \textit{pinned regime} while those above it ($F_d>0.008\epsilon_0$) constitute the \textit{moving regime}.

We have repeated these numerical operations for non-interacting flux lines and found the results to be very similar: $r_g$ once again peaked around $F_d \approx 0.006\epsilon_0$ which is also the value at which $v$ started assuming non-zero values, and $f_p$ began decaying from its steady initial value, indicating that for our purposes, the ranges for the pinned and moving regimes remain essentially unchanged for the non-interacting case.

\section{Results: Relaxation post drive quench}\label{secRel}

\subsection{Quenches within the moving regime}\label{secMovMov}

\begin{figure}
\centering
  \subfloat{\includegraphics[width=.97\linewidth]{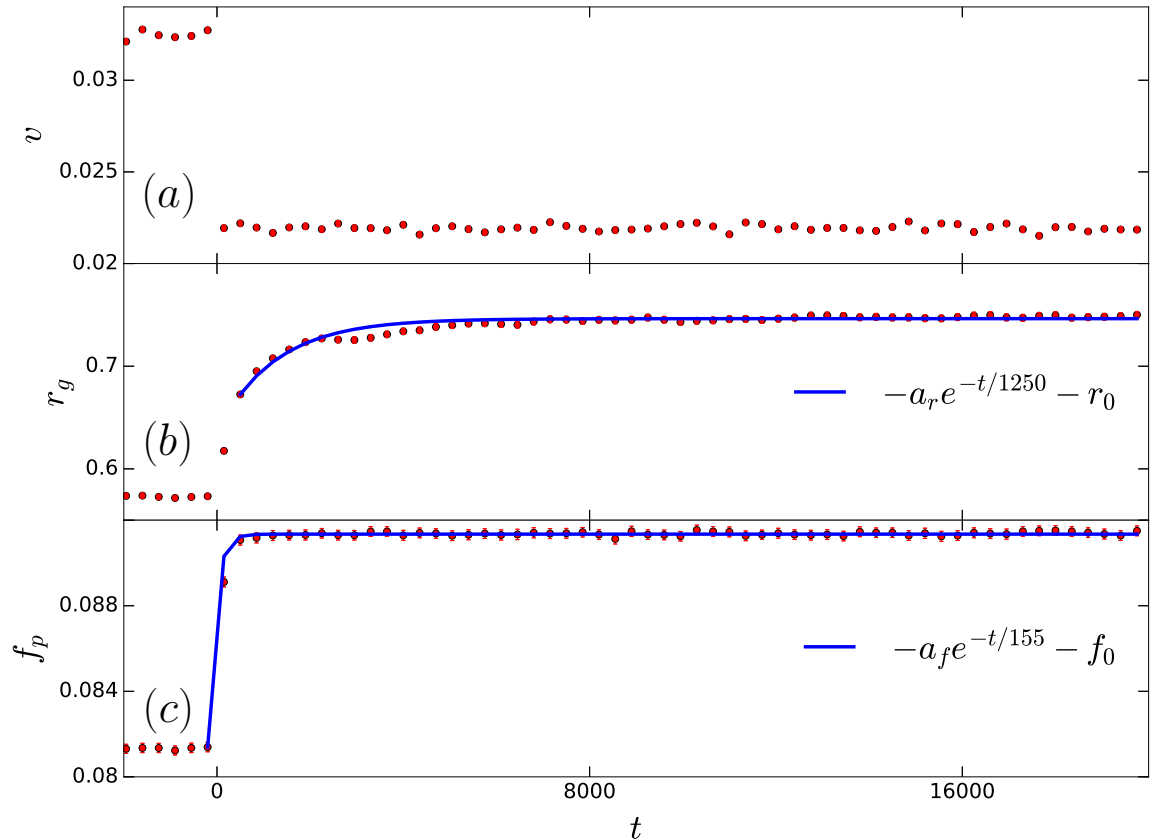}
  \llap{\parbox[b]{5.35in}{{\Large $(a)$}\\\rule{0ex}{3.3in}}}
  \llap{\parbox[b]{5.4in}{{\Large $(b)$}\\\rule{0ex}{2.in}}}
  \llap{\parbox[b]{5.45in}{{\Large $(c)$}\\\rule{0ex}{.6in}}}}
  \caption{Relaxation of the (a) velocity $v$ (units of $b_0/t_0$), (b) radius of gyration $r_g$ (units of $b_0$), and (c) fraction of pinned line elements $f_p$ with time (units of $t_0$) for a system of interacting flux lines in the presence of point-like disorder, following a drive down-quench from $F_d=0.035\epsilon_0$ to $0.025\epsilon_0$ (moving to moving regime), with relaxation times $\uptau_v=0$, $\uptau_{r_g}=1250 t_0$, and $\uptau_{f_p}=155 t_0$, respectively. Data are averaged over 1000 independent realizations. \label{plot_v_r_f_mov}}
\end{figure}

In a first set of numerical experiments, we quench the drive in a moving ($F_d = 0.035\epsilon_0$) steady-state system of vortex lines, in the presence of point-like disorder, to $F_d = 0.025\epsilon_0$, a drive value also in the moving regime.

For interacting lines, upon quenching, the mean velocity $v$ of the lines drops suddenly (Fig. \ref{plot_v_r_f_mov}a) due to the system being in an overdamped Langevin regime which effectively renders the elastic lines massless; the lines have no inertia and an instantaneous change in drive causes an equally abrupt change in velocity. At the moment of quench, the mean radius of gyration of these interacting lines starts growing (Fig. \ref{plot_v_r_f_mov}b). This is in agreement with our expectations: the reduced mean vortex velocity allows easier trapping of the lines by the pins present in the sample. This increased susceptibility to pinning coupled with thermal wandering results in the lines assuming increasingly distorted configurations, whence their roughness is enhanced as a function of time. The growth of $r_g$ is fast (exponential) and stabilizes to a new steady-state value within a relaxation time $\uptau=1250t_0$. This exponential relaxation implies that when quenching within the moving regime, the system transitions from one non-equilibrium steady state to another quickly. The fraction of pinned line elements $f_p$ also grows rapidly (Fig. \ref{plot_v_r_f_mov}c) and reaches a new steady-state value after $\uptau=155t_0$ upon quenching the drive. This is to be expected since the lowered velocity of the lines means that a larger fraction of line elements in the system are susceptible to trapping by the pins. The free parameters for the mathematical functions that have been fitted to the data in Fig. \ref{plot_v_r_f_mov} and subsequent figures were determined using the method of least squares. The time evolution of one-time physical properties $v$, $r_g$, and $f_p$ for non-interacting lines was found to be very similar to that for the interacting case discussed above, with comparable exponential relaxation times ($\uptau_{v}=0$, $\uptau_{r_g}=1120t_0$, and $\uptau_{f_p}=149t_0$). The effect of interactions on flux line dynamics only becomes evident when we study two-time height autocorrelations $C(t,s)$.

\begin{figure}
\centering
  \subfloat{\includegraphics[width=.97\linewidth]{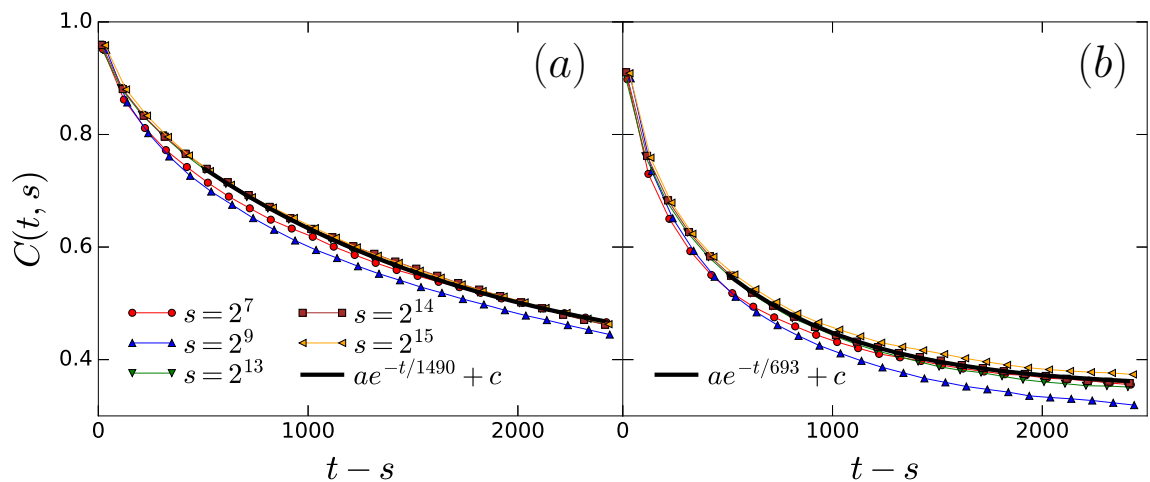}
  \llap{\parbox[b]{3.3in}{{\Large $(a)$}\\\rule{0ex}{2.1in}}}
  \llap{\parbox[b]{0.5in}{{\Large $(b)$}\\\rule{0ex}{2.1in}}}}
  \caption{Evolution of the height autocorrelation function $C(t,s)$ as a function of the post-snapshot time $t-s$ (units of $t_0$), for waiting times $s =$ $2^7t_0$, $2^9t_0$, $2^{13}t_0$, $2^{14}t_0$, and $2^{15}t_0$, for systems of (a) non-interacting and (b) interacting flux lines in the presence of point disorder, following a drive down-quench from $F_d=0.035\epsilon_0$ to $0.025\epsilon_0$ (moving to moving regime). Time translation invariance is obeyed in both cases for larger waiting times ($s\ge2^{13}t_0$), as seen by the collapse of the corresponding $C(t,s)$ curves onto stationary curves that show exponential relaxation, with the interacting lines relaxing faster ($\uptau=693t_0$) than the non-interacting lines ($\uptau=1490t_0$). Data are averaged over 1000 independent realizations. \label{HA_mov}}
\end{figure}

We have measured $C(t,s)$ for waiting times $s =$ $2^{7}t_0$, $2^{9}t_0$, $2^{13}t_0$, $2^{14}t_0$, and $2^{15}t_0$ as a function of time elapsed post-quench $t-s$ (Fig. \ref{HA_mov}). In both, non-interacting (Fig. \ref{HA_mov}a) and interacting (Fig. \ref{HA_mov}b) cases, the autocorrelations for the higher waiting times ($s\geq2^{13}t_0$) are observed to be time-translation invariant, i.e., they coincide and display exponential relaxation, with the interacting lines relaxing faster ($\uptau=693t_0$) than the non-interacting ones ($\uptau=1490t_0$). The faster relaxation in the presence of vortex interactions can be attributed to caging effects. The repulsions force the lines apart, resulting in faster depinning of the line elements and hence straightening of the lines as well as the confining of these straightened lines into a moving lattice. This quick straightening results in the lines becoming spatially uncorrelated with their initial horizontal configurations faster than in the non-interacting case, thus explaining the faster height autocorrelation decay. Time-translation invariance is broken, however, when we go to shorter waiting times ($2^7t_0$, $2^9t_0$) for both the non-interacting and interacting cases. This is to be expected since the waiting times in question are shorter than the relaxation time ($\uptau=693t_0 \approx 2^{9.4}t_0$), a regime where the system has not yet forgotten its initial state; therefore its relaxation behavior is dependent on when we start measuring the autocorrelation function, i.e., it depends on the waiting time $s$. The observation of time translation invariance in the evolution of the height autocorrelation functions corresponding to higher waiting times rules out the possibility of physical aging in the system, as was already hinted at by the exponentially fast relaxation of the radius of gyration $r_g(t)$.

\subsection{Quenches from the moving into the pinned regime}\label{secMovPin}

For our next set of numerical experiments, we quench the drive of a system of flux lines in the moving regime ($F_d = 0.025\epsilon_0$) to $F_d = 0$ in the pinned regime.

\begin{figure}
\centering
  \subfloat{\includegraphics[width=.97\linewidth]{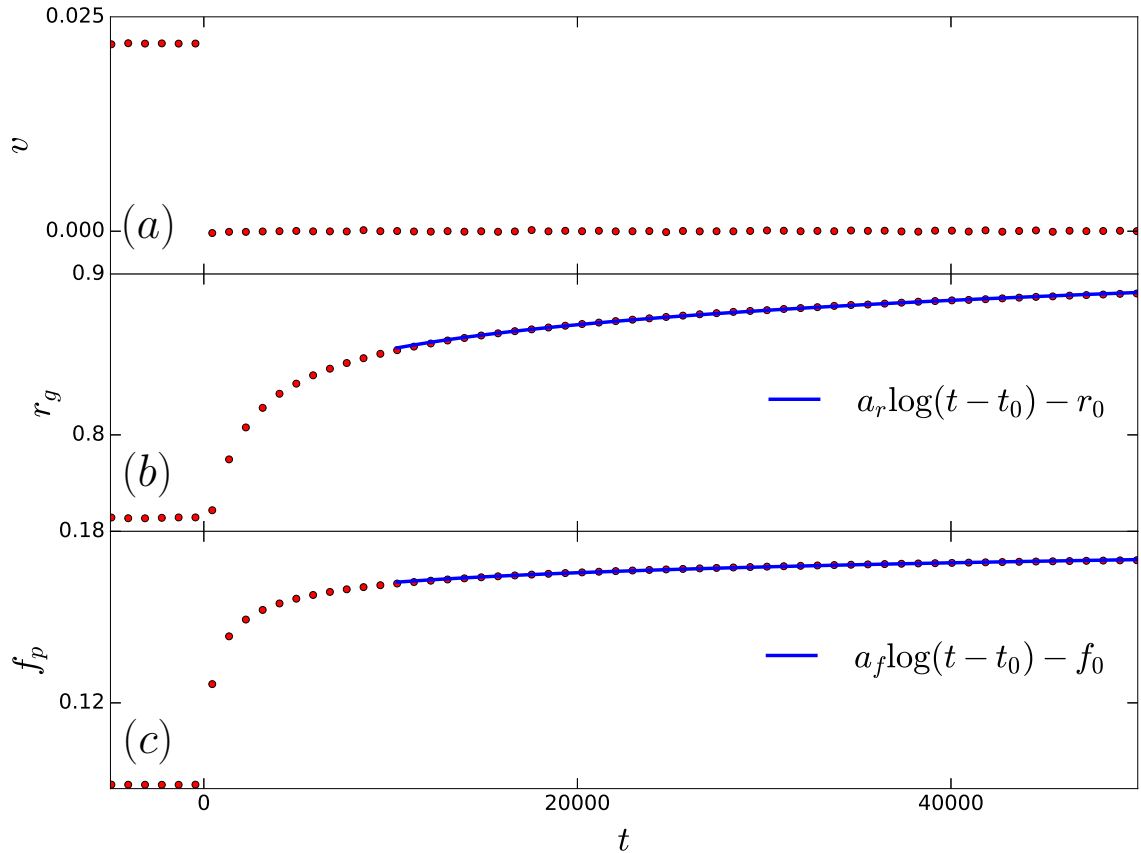}
  \llap{\parbox[b]{5.4in}{{\Large $(a)$}\\\rule{0ex}{3.2in}}}
  \llap{\parbox[b]{5.45in}{{\Large $(b)$}\\\rule{0ex}{1.9in}}}
  \llap{\parbox[b]{5.5in}{{\Large $(c)$}\\\rule{0ex}{.5in}}}}
  \caption{Relaxation of the (a) mean vortex velocity $v$ (units of $b_0/t_0$), (b) radius of gyration $r_g$ (units of $b_0$), and (c) fraction of pinned line elements $f_p$ with time $t$ (units of $t_0$) for a system of interacting flux lines in the presence of point-like disorder, following a drive down-quench from $F_d=0.025\epsilon_0$ to $0$ (moving to pinned regime). $v$ drops instantaneously, while both $r_g$ and $f_p$ relax logarithmically slowly ($a_r=0.05b_0$, $a_f=0.01$) with $t$. Data are averaged over 1000 independent realizations. \label{plot_v_r_f_pin}}
\end{figure}

\begin{figure}
  \centering
  \subfloat{\includegraphics[width=.47\linewidth]{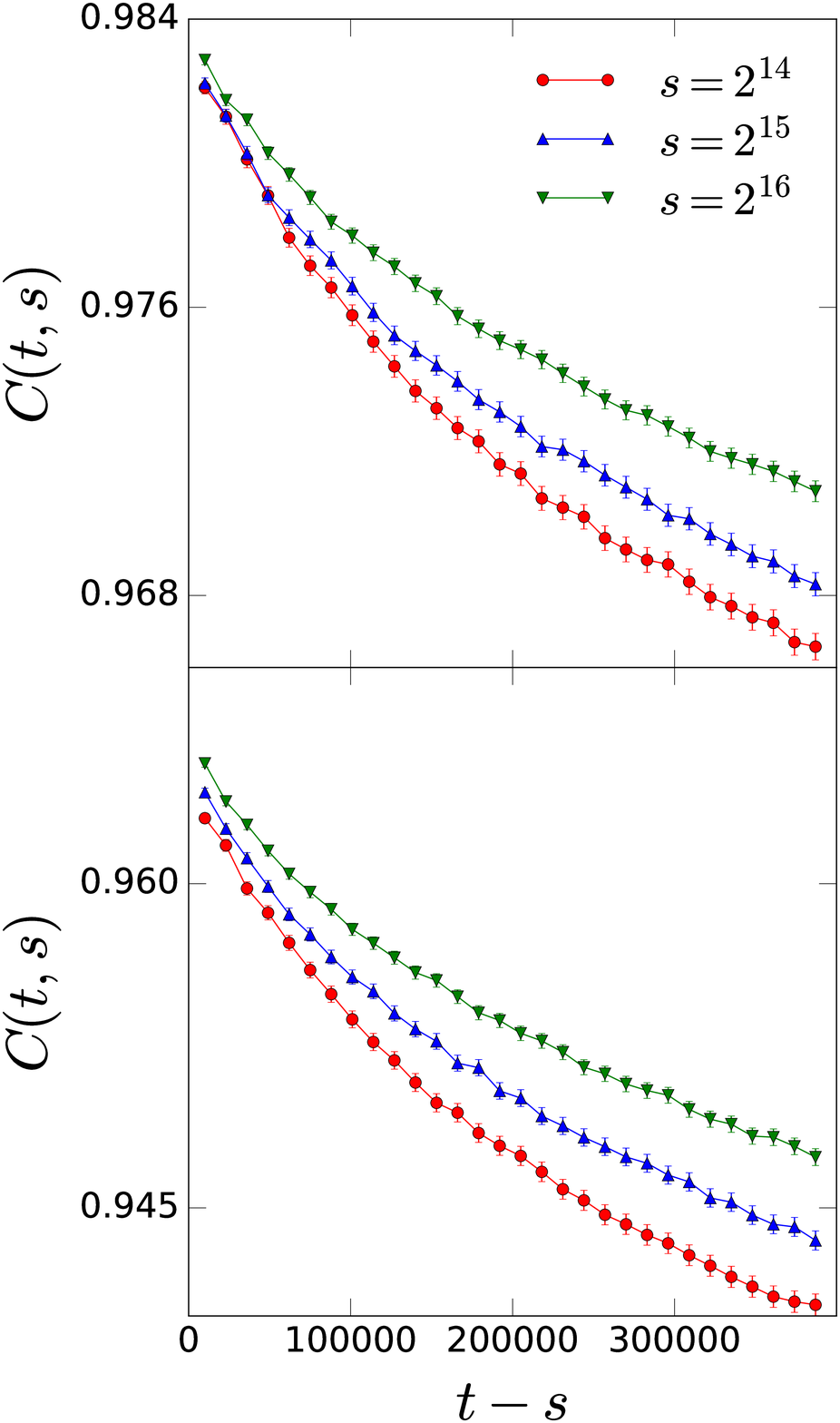}
  \llap{\parbox[b]{2.2in}{{\Large $(a)$}\\\rule{0ex}{2.85in}}}
  \llap{\parbox[b]{2.25in}{{\Large $(c)$}\\\rule{0ex}{.5in}}}} 
  \subfloat{\includegraphics[width=.485\linewidth]{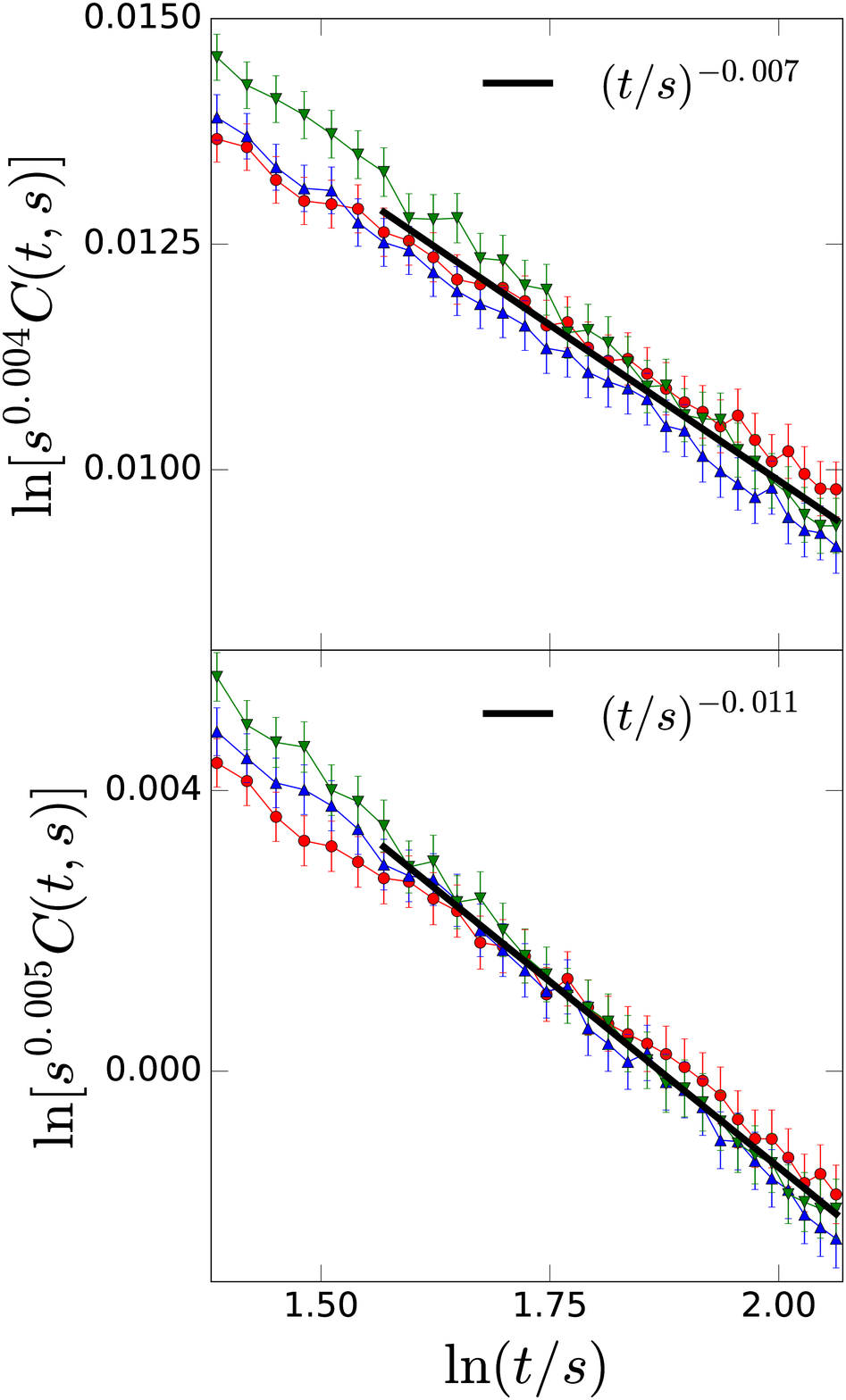}
  \llap{\parbox[b]{2.2in}{{\Large $(b)$}\\\rule{0ex}{2.85in}}}
  \llap{\parbox[b]{2.25in}{{\Large $(d)$}\\\rule{0ex}{.5in}}}}
\caption{Height autocorrelation function $C(t,s)$ as a function of $t-s$ (a, c), and scaled height autocorrelation $s^{-b}C(t,s)$ as a function of $t/s$ (b, d), for systems of (a, b) non-interacting and (c, d) interacting flux lines in the presence of point disorder, following a drive down-quench from $F_d=0.025\epsilon_0$ to $0\epsilon_0$ (moving to pinned regime). Time translation invariance is broken (a, c) and dynamical scaling is observed (b, d) in both cases, with scaling exponents $(b$, $\lambda_C/z)$ found to be $(0.004$, $0.006)$ and $(0.005$, $0.011)$, respectively, for the (b) non-interacting and (d) interacting cases. Data are averaged over $10,000$ independent realizations. \label{HA_pin}}
\end{figure}

For the interacting lines, at the moment of quench, the velocity $v$, once again as in the case of quenches within the moving regime, drops instantaneously to zero (Fig. \ref{plot_v_r_f_pin}a) as the system enters a pinned state. The drop in velocity is accompanied by growth of the radius of gyration $r_g$ (Fig. \ref{plot_v_r_f_pin}b). This growth is very slow, however, when compared to the exponentially fast relaxation of the radius of gyration that we observed in the case of quenches within the moving regime. Here, the relaxation is slow enough that the radius of gyration cannot stabilize to a steady value on the time scales we are exploring, and instead shows a logarithmic growth with time. Initial attempts to fit the $r_g$ data to a power law by the method of least squares yielded exponents quite close to zero. A logarithmic function was therefore tested and found to provide a superior fit (smaller residuals) to the data than any temporal power law. This slower logarithmic growth can be attributed to the system entering a Bragg glass phase where the system of flux lines has access to many metastable states, each corresponding to a unique configuration. These states have a negligible mean velocity but have similar probabilities associated with several different pinning configurations. For interacting lines, the growth in $r_g$ does not persist indefinitely, but terminates at a certain upper value of time $t$. This is a consequence of the caging effect of the repulsive vortex interactions on the growth of the time-dependent correlation length $L(t)$ associated with the flux lines \cite{Pleimling2015}. However this caging effect is not yet perceptible in the data shown in Fig. \ref{plot_v_r_f_pin}. The interaction-induced caging effect will also affect the behavior of the two-time height autocorrelation functions at very long times. Another one-time quantity that displays slow logarithmic growth post quench as the system enters the glassy pinned state is the fraction of pinned line elements $f_p$  (Fig. \ref{plot_v_r_f_pin}c), in contrast to the fast exponential growth and stabilization of the quantity seen for quenches within the moving regime (moving-to-moving quenches). For the relaxation of one-time quantities $v$ ($\uptau = 0$), $r_g$ ($a_r = 0.06b_0$), and $f_p$ ($a_f = 0.01$) in the interaction-free situation, as in the case of moving-to-moving quenches, we did not find remarkable qualitative differences compared to the system with interacting lines.

The two-time height autocorrelations $C(t,s)$ for quenches into the pinned regime display slow temporal relaxation accompanied by the breaking of time translation invariance for both non-interacting (Fig. \ref{HA_pin}a) and interacting lines (Fig. \ref{HA_pin}c). This is in contrast to the situation for quenches within the moving regime, where time translation invariance was clearly observed for the entire period of measurement for waiting times greater than the relaxation time of the system. We checked the autocorrelations for dynamical scaling by testing a range of scaling exponents $b$ in the following way. For each $b$ under consideration, we plotted the three $s^bC(t,s)$ curves ($s = 2^{14}t_0$, $2^{15}t_0$ and $2^{16}t_0$) against $t/s$. We then employed a least-squares algorithm to compare these functions and identified the value of $b$ that rendered the best dynamical scaling collapse. For the non-interacting (Fig. \ref{HA_pin}b) and interacting (Fig. \ref{HA_pin}d) cases, the algorithm yielded pairs of dynamical aging scaling exponents $(b$, $\lambda_C/z) = (0.004$, $0.007)$  and $(0.005$, $0.011)$, respectively, for which the individual height autocorrelation curves collapsed onto a master curve, a clear indication of physical aging in the system. The scaling only emerges for larger $t/s$, when the system has had sufficient time to overcome the initial large fluctuations that immediately follow the quench, and to enter the aging scaling regime. 

For interacting lines, the aging scaling regime will be cut short at very long times by the caging effect of the repulsive vortex interactions (also responsible for limiting the growth of $r_g$) \cite{Pleimling2015}. The scaling form for simple aging given in \eref{scaform} is a special case of the more general scaling form $f_C(t,s)\sim[L(t)/L(s)]^{-\lambda_C}$. The simple aging form arises from the general case when $L(t)$ grows as a simple power law of $t$. The algebraic growth $L(t)\sim t^{1/z}$ with the dynamic scaling exponent $z$ is limited by the interaction-induced caging effect. The aging scaling exponents $b$ seen here are over an order of magnitude smaller than those obtained in previous studies: one on the aging of randomly placed, interacting flux lines in the absence of drive \cite{Dobramysl2013} and another on relaxation following temperature and magnetic field quenches, also for randomly placed flux lines without drive \cite{Assi2015}. In the case of drive quenches as presented here, we have verified that during the initial pre-quench, high-drive ($F_d = 0.035$) period of the simulation, the flux lines constitute a highly correlated moving lattice. This is in contrast to the previous studies where, on account of the absence of drive, the initial disorder dominated state was always random and uncorrelated. We can thus infer that the initial conditions have a significant influence on the aging scaling exponents, with a correlated initial state yielding far smaller aging scaling exponents compared to an uncorrelated one.

\section{Conclusion} \label{conclusion}
In this paper, we have investigated the long-time relaxation features of driven magnetic flux vortices in type-II superconductors following sudden quenches of external current. In order to study the post-quench dynamics of these vortices in the presence of uncorrelated point-like disorder, we modeled them as directed elastic lines in the presence of localized pinning centers, and solved the associated Langevin molecular dynamics equations numerically. In the simulations we maintained a constant ambient temperature. The external current quenches were realized in the form of instantaneous changes in the drive, a quantity in the elastic line model that mimics the Lorentz force exerted by external current on the flux vortices. In this study, we focused on two types of drive quenches, those within the moving regime and those from the moving regime into the pinned regime.

For quenches within the moving regime, we have studied the effects of the vortex-vortex repulsive interactions on the relaxation kinetics of the vortices by performing drive quenches in the system with the interactions initially absent or in effect. In both cases, drive quenches within the moving phase result in fast exponential relaxation of the system from one non-equilibrium steady state to another, as evidenced by the rapid temporal evolution of one-time observables such as the mean radius of gyration of the lines and the fraction of pinned line elements. The two-time height autocorrelation functions for different waiting times display similar fast exponential relaxation as the one-time quantities, along with time translation invariance, firmly eliminating the possibility of physical aging in the case of quenches within the moving regime. When turned on, the screened logarithmic repulsive interactions between the flux lines significantly speed up the exponential relaxation of the height autocorrelations with the associated relaxation time being around half that for quenches with no interactions present.

For our study on drive quenches from the moving to the pinned regime, in stark contrast to quenches within the moving regime, the relaxation of the system after the quench is much slower, which is seen in the non-exponential, logarithmic time evolution of the radius of gyration and fraction of pinned line elements. This indicates that the system fails to reach a steady state when quenched into the pinned regime on time scales that are on the order of the simulation duration. The two-time height autocorrelations show breaking of time translation invariance, accompanied by dynamical scaling with $t/s$, evidence for aging in the system, as we quench it from a moving non-equilibrium steady state into a pinned, glassy one. The $t/s$ range for which simple aging is applicable for interacting lines is bound by the limiting of the algebraic growth of the characteristic time-dependent correlation length $L(t)$, a consequence of the caging of the flux lines by the repulsive vortex interactions. Correlated initial conditions as with the moving lattice seen in the initial state in our study yield markedly smaller aging scaling exponents compared to uncorrelated initial conditions such as those obtained in previous investigations where the flux lines were initially randomly distributed.

\section*{Acknowledgments} \label{acknowledgements}
This research is supported by the U.S. Department of Energy, Office of Basic Energy Sciences, Division of Materials Sciences and Engineering under Award DE-FG02-09ER46613.

\section*{References}
\bibliographystyle{iopart-num}
\bibliography{bibliography}

\end{document}